\documentclass{webofc}
\usepackage[varg]{txfonts}   
\begin{document}
\title{Addressing Scalability with Message Queues: Architecture and Use Cases for DIRAC Interware}
%
%

\def\lhcb   {\mbox{LHCb}\xspace}
\def\belle  {\mbox{Belle2}\xspace}
\def\clic   {\mbox{the Linear Collider}\xspace}
\def\lhc    {\mbox{LHC}\xspace}

\def\dirac      {\mbox{\textsc{DIRAC}}\xspace}
\def\lhcbdirac  {\mbox{\textsc{LHCB\dirac}}\xspace}

\author{
  \firstname{Wojciech} \lastname{Krzemien}\inst{1}\fnsep\thanks{\email{wojciech.krzemien@ncbj.gov.pl}} \fnsep\thanks{On behalf of the LHCb collaboration.}
    \and
  \firstname{Federico} \lastname{Stagni}\inst{2} \and 
  \firstname{Christophe} \lastname{Haen}\inst{2} \and
  \firstname{Zoltan} \lastname{Mathe}\inst{2} \and
  \firstname{Andrew } \lastname{McNab}\inst{3} \and
  \firstname{Milosz} \lastname{Zdybal}\inst{4} 
}

\institute{High Energy Physics Division, National Centre for Nuclear Research, PL-05-400 Otwock, Swierk, Poland
\and
         CERN, EP Department, European Organization for Nuclear Research, Switzerland  
\and
         The University of Manchester, Oxford Road, Manchester, M13 9PL, UK
\and
         Institute of Nuclear Physics PAN, Krakow, Poland
 }

\abstract{
The Message Queue (MQ) architecture is an asynchronous communication scheme that provides an attractive solution for certain scenarios in a distributed computing model. The introduction of MQ as an intermediate component in-between the interacting processes allows to decouple the end-points making the system more flexible and providing high scalability and redundancy.

\dirac is a general-purpose interware software for distributed computing systems, which offers a common interface to a number of heterogeneous providers and guarantees transparent and reliable usage of the resources. The \dirac platform has been adapted by several scientific projects, including High Energy Physics communities like \lhcb, \clic and \belle.

A Message Queue generic interface has been incorporated into the \dirac framework to help solving the scalability challenges that must be addressed during \lhc Run3, starting in 2021. It allows to use the MQ scheme for a message exchange among the DIRAC components or to communicate with third-party services. Within this contribution we describe the integration of MQ systems with DIRAC and several use cases are shown. Message Queues are foreseen to be used in the pilot logging  system,  and as a backbone of the DIRAC component logging system and monitoring.
}
\maketitle
\section{Introduction}
\label{intro}
 
We live in a world of large data streams, which are constantly provided  by various sources and need to be processed efficiently. This massive amount of data requires the use of all available processing resources together with an efficient computing model, which is scalable and reliable. 

High Energy Physics (HEP) communities face similar challenges, since the data produced by the experiments' detectors and by the Monte Carlo simulation jobs form a significant data stream that must be processed in a coordinated manner~\cite{LHCb-dirac-upgrade}. For this purpose, several approaches have been proposed,  among them the \dirac framework~\cite{DIRAC-SW,DIRAC-CHEP2012}. \dirac, the interware, is an open-source software platform that provides the interface between the end-user and the underlying resources.

\dirac implements a flexible distributed agent model that assures scalable processing over heterogeneous environments.  
The \dirac interware was adopted as a computing solution by the HEP experiments like \lhcb and \belle, and also by many other projects which use it as a platform to perform advanced GRID operations.

Message Queue (MQ) architectures implement asynchronous communication schemes which fit very well to distributed models based on microservices and provide several advantages, including good scalability and performance. This paper is dedicated to the incorporation of existing MQ solutions in \dirac.

The paper is organized as follows: First, in section~\ref{sec:mq}, the concept of Message Queue communication model is introduced. Section~\ref{sec:dirac-mq} explains the details of Message Queue module implementation into the \dirac framework. Section~\ref{sec:use-cases} presents several use cases. Finally, a summary is given in section~\ref{sec:summary}.

\section{Message Queue Concepts}\label{sec:mq}

\begin{figure*}
\centering
\includegraphics[width=10cm,clip]{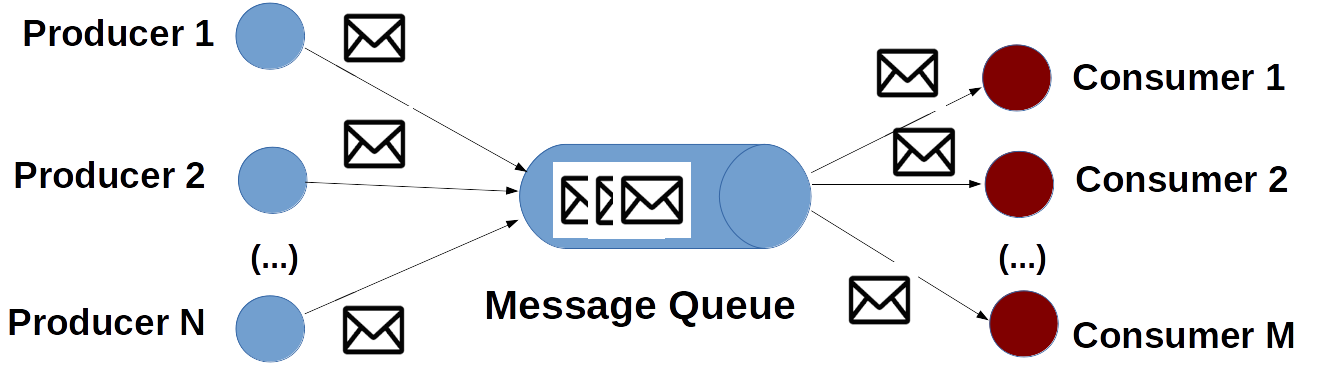}
\vspace*{1cm}       
\caption{Message-queueing asynchronous communication model. Producers send messages that are stored by the intermediate component (Message Queue). The messages can be retrieved by consumers. The Message Queue decouples interacting entities.}
\label{fig-MQscheme}       
\end{figure*}

MQ communication is based on the idea of introducing an intermediate component (queue) in-between interacting entities, which are typically called consumer and producer (see Fig.~\ref{fig-MQscheme}).
A queue acts as a buffer that stores messages sent by producers. This separation allows the communication to become asynchronous as the consumer and producer do not need to interact at the same time. This approach has several advantages: it allows to decouple the end-points making the system more flexible and providing high scalability and redundancy. In some MQ systems additional mechanisms are implemented to ensure, e.g., resilience or message delivery guarantee.  Also, MQ architecture introduces flexibility at the technology level, permitting to interconnect heterogeneous environments.

The MQ paradigm is applicable at very different levels. It may serve as an inter-process communication mechanism acting within one operating system as well as a way of connecting the processes or services in distributed computing models. Various message-oriented middlewares (brokers) have been developed. Among them, open-source solutions such as RabbitMQ~\cite{rabbitmq}, ActiveMQ~\cite{activemq} or Kafka~\cite{kafka} are proven technologies widely used nowadays.

\section{Message Queues in DIRAC}\label{sec:dirac-mq}

\begin{figure*}
\centering
\includegraphics[width=10cm,clip]{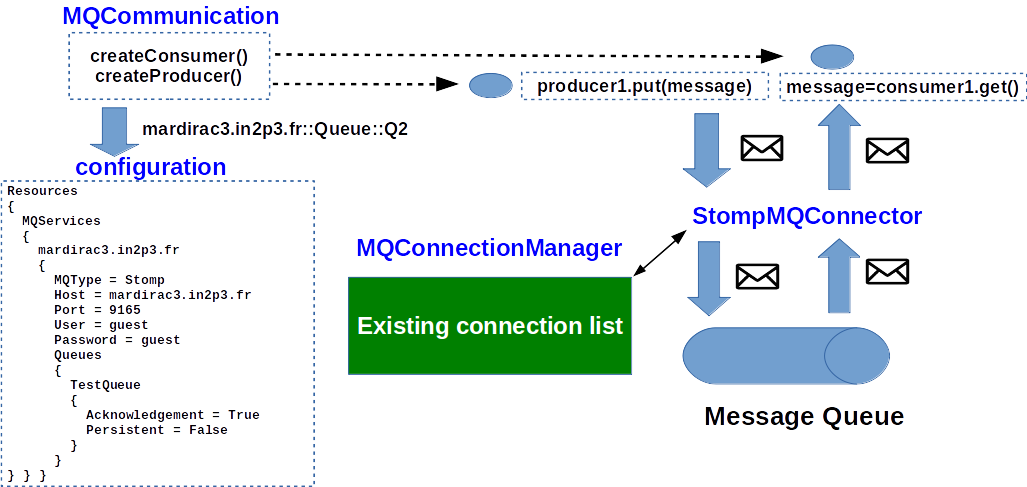}
\vspace*{1cm}       

\caption{In DIRAC, the MQ configuration is defined as resource sections set through a DIRAC Configuration Service (not in the figure). Each section is uniquely identified by the pseudo-url string, e.g., ``mardirac3.in2p3.fr::Queue::Q2'', which can be provided as an argument to the factory methods createConsumer() and createProducer() responsible for the creation of producer and consumer instances. To improve the performance, the same connections can be reused by several consumers or producer. This functionality is provided by MQConnectionManager module. The purpose of MQConnector interfaces is to provide a mechanism that accommodates various communication protocols e.g. STOMP~\cite{stomp}. More details are given in the text.}
\label{mq-architecture}       
\end{figure*}

A generic MQ interface has been introduced in DIRAC version 6, release 17. It allows to connect \dirac components to (external) MQ services and to exchange messages with them. An access to the MQ services is realised via logical Queues or Topics~\cite{patterns}. The architecture of the MQ interface is presented in Fig.~\ref{mq-architecture}.

The MQCommunication interface provides factory methods to create MQConsumer and MQProducer instances which can be used to exchange messages with the MQ. 
The only requirement for the message format is that it must conform with a {\it json} structure.
The configuration settings are loaded from the \dirac Configuration Service, identified by the pseudo-url string, e.g., ``mardirac3.in2p3.fr::Queue::Q2'', which is provided as an argument to the factory methods.
MQConnectionManager manages internally the list of open connections and assures thread-safe access. The same connections can be reused by several consumers/producers to improve the performance. 
The specialisation of the MQConnector then provides an interface mechanism tailored to a chosen MQ communication protocol.  
Currently, the handler implementation for Simple Text Orientated Messaging Protocol (STOMP)~\cite{stomp} standard is available. Both user-password and SSL certificates authentication mechanisms are supported. 
The implementation was tested with two message brokers: RabbitMQ~\cite{rabbitmq} and ActiveMQ~\cite{activemq}.
The existing scheme can be easily extended by adding a specialized module, e.g., to support more MQ protocol types.  

\section{Use Cases}
\label{sec:use-cases}
In this section we briefly describe several use cases in which the MQ components are being used.

The MQ has been used as part of the perfSONAR-\dirac bridge architecture  that is used for network performance monitoring, providing the metrics, and for network problem identification. More details can be found in~\cite{perfsonar, McKee_2017}.

\begin{figure*}
\centering
\includegraphics[width=10cm,clip]{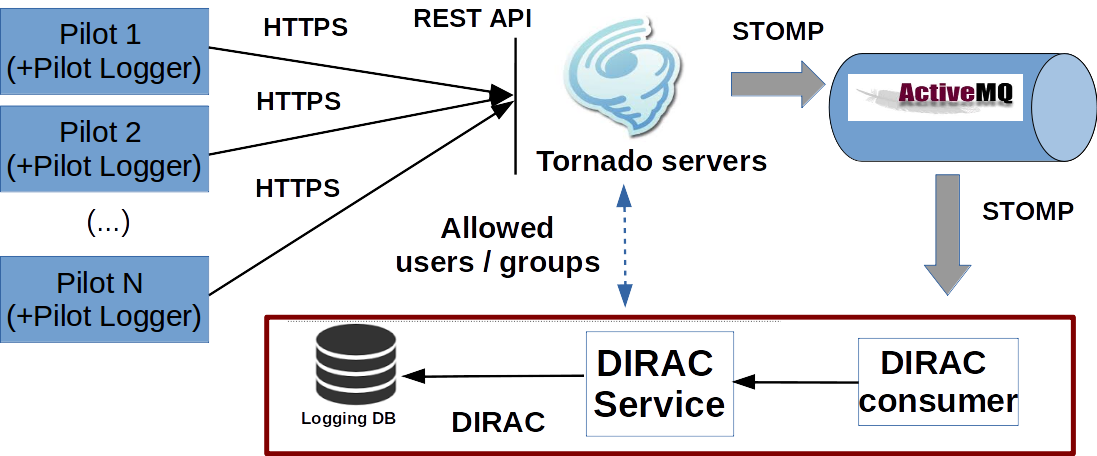}
\vspace*{1cm}       
\caption{Architecture for the universal pilot logging system. The MQ broker (in the scheme ActiveMQ~\cite{activemq}) collects the log information provided by the pilot. The transfer is performed via the Tornado server~\cite{tornado}, that assures both authentication and authorization. The MQ consumers receive the logs that are finally stored by a dedicated DIRAC service.}
\label{fig-pilotLogging}       
\end{figure*}

The \dirac system is installed on worker nodes (WN) by distributed agents called pilots~\cite{DIRAC-pilots-2015,Stagni_2017}. The development of a universal and scalable logging system for all pilots is also foreseen to accommodate the use of the MQ (see Fig.~\ref{fig-pilotLogging}). Due to the variability of WN types, it is challenging to provide a generic solution that would provide information about possible failures during, e.g., the installation or configuration phases. The proposed architecture is shown in Fig.~\ref{fig-pilotLogging}. The Pilot Loggers are components of the new \dirac  pilot generation. They are responsible for sending the logs to a dedicated system. The development is ongoing. 

MQ is also used as the main buffer for internal DIRAC services` logging systems. This system is currently in production used together with the CERN ActiveMQ system. Finally, MQ is implemented as a failover mechanism for the ElasticSearch~\cite{elastic} in \dirac  monitoring services~\cite{Mathe_2015}. The monitoring system is dedicated to monitoring various components of DIRAC. 
It is based on Elasticsearch distributed search and a NoSQL analytics database. The implemented failover mechanism allows to redirect the logs to a dedicated MQ server. This solution has been tested with the RabbitMQ server. 

\section{Summary}
\label{sec:summary}
The MQ generic interface has been successfully introduced in the \dirac framework.
It is being used as an important part of the \dirac service logging system, as a failover mechanism for the \dirac Monitoring System, and it is foreseen to play an important role in the universal pilot logging architecture being developed.


\end{document}